\documentclass{elsart}

\usepackage{amssymb}
\usepackage{epsf,colordvi,color,amsbsy}
\usepackage[dvips]{graphicx}
\begin{document}

\begin{frontmatter}

 \title{Solar cycle activity: an early prediction for cycle \#25}

\author[]{Stefano Sello \corauthref{}}

\corauth[]{stefano.sello@enel.com}

\address{Mathematical and Physical Models, Enel Research, Pisa - Italy}

\begin{abstract}

Solar activity forecasting is an important topic for numerous
scientific and technological areas, such as space mission operations,
electric power transmission lines, power transformation stations and earth geophysical and climatic impact.
Nevertheless, the well-known difficulty is how to accurately
predict, on the basis of various recorded solar activity indices,
the complete evolution of future solar cycles, due to highly
complex dynamical and stochastic processes involved, mainly related to
interaction of different components of internal magnetic fields.
There are two main distinct classes of solar cycle prediction
 methods: the precursor-like ones and the mathematical-numerical ones.
  The main  characteristic of precursor techniques, both purely solar and
geomagnetic, is their physical basis. Conversely, the non-precursor methods
use different mathematical and/or numerical properties of the known temporal
evolution of solar activity indices to extract useful information
for predicting future activity. For current solar cycle \#24
we obtained fairly good statistical performances from both precursor and
purely numerical methods, such as the so-called solar precursor
and nonlinear ones. To further check the performances of these
prediction techniques, we compared the early predictions for the
next solar cycle \#25. Preliminary results support some coherence
of the prediction methods considered and confirm the current trend
of a relatively low solar activity.

\end{abstract}
\end{frontmatter}

 \section{Introduction}

The current solar cycle \#24 has now completed its declining phase reaching the minimum state around September 2018, following a
multipeaked maximum around 2012.20 (lower peak:
$R_z=98$) and 2014.31 (main peak: $R_z=115.6$), with a prominent Gnevyshev gap, for
the monthly smoothed sunspot numbers, as revised by
SIDC from 1 July 2015. This new version of $R_z$ is used here. Note that it is $40\%$ - $70\%$ larger than the previous older version (WDC-SILSO
\cite{SIDC}). As reviewed by Pesnell (\cite{Pesnell16}), many solar cycle forecasting methods predicted
higher values, including statistical, curve fitting, spectral one
and some geomagnetic precursor-like ones. 
It is important to note that except from few examples, most of the predictions predict only the amplitude and time of the next solar maximum. Missing
is the detailed shape of the curve time behavior, including the timing of cycle extrema, which are usually found by examining
not only the sunspot number record but also other measures of solar activity, such as the flare rate or
magnetic flux, that may peak at different times of the sunspot cycle. This is a more difficult task than the maximum amplitude prediction as well shown in Sello (\cite{Sello16}).
Those methods using multiple information sources, such as the precursors, have smaller average errors and this led to the general conclusion that precursors were superior, which seems to be valid also for Solar Cycle \#24, which was considered to be an atypical sunspot cycle. In fact, some predictions indicated very high levels of activity, others very low even suggesting the tendency to a grand minimum of solar activity. The observations show a below-average amplitude sunspot cycle, with all the usual activity present at quite reduced amplitudes, see for details: Pesnell, (\cite{Pesnell16}).

As Solar Cycle \#24 fades and reaches its minimum phase, the anticipation of Solar Cycle \#25 begins. Several predictions of the amplitude of Solar Cycle \#25 have already appeared, along with more difficult longer-term forecasts of solar activity.
 As few examples we cite here Helal and Galal [2013] which used a correlation between the number of spotless days and the upcoming solar maximum to estimate that Solar Cycle \#25 will have: $R_z=118.2$, peaking $4.0$ years after the solar minimum. Yoshida [2014] used correlations between $R_z$  before minimum with the upcoming solar maximum, using the symmetries of the even/odd cycles to derive the prediction: $R_z = 115.4 \pm 11.9$.  Cameron,  Jiang, and Schüssler (2016) using surface flux transport simulations for the descending phase of Cycle \#24 with random sources (emerging bipolar magnetic regions) provide a prediction of the axial dipole moment during the incoming activity minimum. The empirical correlation between the dipole moment during solar minimum and the strength of the subsequent cycle suggests that Cycle \#25 will be of moderate amplitude, not much higher than that of the current cycle. The authors stress that the intrinsic uncertainty of such predictions resulting from the random scatter of the source properties is considerable and limits the reliability with which such predictions can be made before activity minimum is reached, as we will note below also for our nonlinear method. Among predictors class one of the most efficient and the first physical based precursor is the Solar Polar Field Precursor Method developed by Schatten et al.  (\cite{Schatten78}), (\cite{Schatten08}) and it is based on polar fields intensity near the minimum as solar activity precursor.

Longer-term predictions are more qualitative and derive an envelope of activity until 2035 [Shepherd et al., 2014]. Their predictions tend to suggest smaller amplitudes of solar cycles \#25 and \#26, as the envelope of activity continues the decline of the last four sunspot cycles.
Here, we refined our nonlinear prediction method to produce a fairly reliable early prediction for the full shape curve of the next solar cycle \#25, see: Sello, (\cite{Sello01}, \cite{Sello03} ).
Previous analyses clearly showed that this prediction method is able to give a sufficiently reliable prediction only after the time of the previous cycle minimum, Sello, (\cite{Sello12}). This is mainly due to the fact that using the maximum amplitudes and the declining phase only of the current cycle, there is not sufficient information about the amplitude and timing of the incoming minimum phase. Thus, a too early prediction tends to underestimate the deep and timing of minimum with a premature increase of the next solar cycle often with an overestimation of the peak amplitude. This behavior is mitigated in our refined prediction method, using a higher value for the embedding dimension. In the following sections we recall briefly the general idea of the method and we show the current prediction for cycle \#25 after a calibration of its parameters based on the best prediction obtained for previous cycle \#24.

\section{Nonlinear dynamics method}

The intrinsic complexity in the temporal behavior of sunspot
numbers (irregular and intermittent) and other solar activity
indices suggested the possibility of a nonlinear (sometimes
chaotic) dynamics governing the related physical processes. In
previous works we have shown the results of a numerical method
based on nonlinear dynamics techniques, properly designed to
predict the medium-time evolution of solar cycles
(Sello \cite{Sello01},\cite{Sello03},\cite{Sello12} ). In particular, the prediction model is
based on the assumption that the underlying dynamics, driven by
the evolution of magnetic fields during a solar cycle, is well
described by a nonlinear deterministic behavior in
 a proper embedding space, i.e. a space that mimics the original phase space, (Abarbanel et al. \cite{Abarbanel90}):

  \begin{equation} \label{eq:map}
  f^T(\vec{y_t})=\vec{y}_{t+T}
  \end{equation}

for a given embedding vector: $\vec{y}$. The inverse problem to
be solved consists of the computation of the smooth map $f^T$,
given a scalar time series (here the sunspot numbers). Following
the approach given by Zhang (\cite{Zhang96}), we designed a
modified improved version that uses different tools of nonlinear
dynamics theory, such as mutual information, linear and nonlinear
redundancy, Lyapunov's spectrum etc. For more detailed information about this nonlinear approach to solar cycle prediction the reader can refer to Sello, (\cite{Sello01}).

\section{An early prediction of Cycle \#25}

A proper refinement of the above nonlinear method, using extended  parametric simulations,  indicates that describing the dynamic evolution in a 7-dimensional embedding space, with a time lag for the embedding vectors sufficient to produce a little entanglement of the vectors in the phase space, it is able to better predict the behavior of the solar cycle curve for the next solar cycle \#25 starting with the information available at or just after the minimum.  In particular, we can predict the multipeaked structure of the cycle, a difficult task for all prediction methods.

The prediction of the next solar cycle \#25 is given using an optimal set of nonlinear embedding parameters. As noted by Sarp et al. (2018), other parameters are important, as the starting point, for reliable predictions of the nonlinear method. The known data used are from WDC-SILSO, Royal Observatory of Belgium, Brussels, up to January 2019 where the minimum is well established.

The predicted peak amplitude for the monthly smoothed sunspot numbers in the next solar cycle \#25
is near $107 \pm 10$, peaking around July $2023 \pm 1$ (see fig.1).  This prediction confirms the current trend towards progressive reduced solar activity cycles.

\begin{figure}
\resizebox{\hsize}{!}{\includegraphics{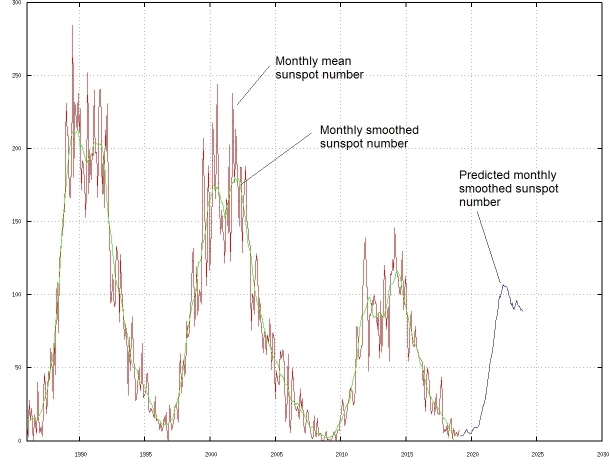}}
 \caption{Predicted evolution for the solar cycle \#25 using the information available up to January 2019 and a refined version of the nonlinear method.  Red: monthly mean sunspot number (WDC-SILSO), Green: monthly smoothed sunspot number (WDC-SILSO), Blue: Predicted monthly smoothed sunspot number. }
 \label{lab1}
\end{figure}

Taking into account the uncertainties, our nonlinear prediction method appears quite coherent, both in amplitude and in phase,
with other different prediction methods; further, many of these early predictions for the next solar cycle \#25, support the tendency towards a
reduction in the intensity of solar cycle activity. Here we cite, as few examples, the work by Singh et al. (2019) where, applying different techniques of time series analysis (Hodrick Prescott filtering method, Hurst exponent, etc.) on various solar activity parameters (sunspot numbers, F10.7 cm index and Lyman alpha index), they found a predicted maximum value of sunspot number as $89 \pm 9$ at February 2024; the work by Gopalswamy, et al. (2018) where, using microwave imaging observations from the Nobeyama Radioheliograph at 17 GHz for long-term studies of solar activity, they found that the smoothed sunspot numbers in the southern and northern hemispheres can be predicted as 89 and 59, respectively, indicating that cycle 25 will not be too different from cycle 24 in its strength; the work by Tan, Baolin (2019), using statistical correlations in the shape of previous solar cycles, found that the Cycle 25 is inferred possibly to be a weak cycle. However, there are also notable exceptions with significant differences in the predicted amplitude and phase of the maximum peak for the next solar cycle: in the cited work by Sarp et al. the authors, using a similar nonlinear dynamics method, predicted that the maximum of Solar Cycle 25 will be at the year $2023.2 \pm 1.1$ with a peak sunspot number of $154 \pm 12$, thus significantly higher that the previous solar cycle; the work by Li, et al. (2018), using the relations among the feature parameters of solar cycles under the bimodal distribution for the modern era cycles (10-23), they predict the maximum amplitude of solar cycle 25 of $168.5 \pm 16.3$ in October 2024  thus quite stronger than solar cycle 24; in the recent work by Pesnell and Schatten, (2018), using the well known "Solar Dynamo" (SODA) Index which combines values of the solar polar magnetic field and the solar spectral irradiance at 10.7 cm to create a precursor of future solar activity, the authors predict a peak activity of about $140 \pm 30$ solar flux units for the 10.7 cm radio flux and a Version 2 sunspot number of $135 \pm 25$, suggesting that Solar Cycle 25 will be quite comparable (but higher) to Solar Cycle 24. The estimated peak is expected to occur near $2025.2 \pm 1.5$ year. Moreover, the authors noted that: "Because the current approach uses data prior to solar minimum, these estimates may improve as the upcoming solar minimum draws closer." Future information with more data will tell us which of these predictions will turn out to be the most correct and reliable. 
%

   \section{Conclusions}

The predictions of atypical solar cycle \#24 activity
show that in general the complexity of the solar processes related to the
evolution of magnetic fields prevents any
 accurate forecasting of the solar cycle full curve shape and in particular an accurate
  estimation for both amplitude (multipeaked) and phase of future cycles.
Nevertheless, different methods, both precursor and
non-precursor, have obtained fairly good overall performances,
suggesting some degree of reliability and usefulness of these prediction
approaches. To further check the performances of some prediction
techniques, here we computed an early full prediction, for the next solar
cycle \#25, using a revised nonlinear dynamics method. Our preliminary result, suggests a maximum peak of about $107 \pm 10$, peaking around July $2023 \pm 1$, supporting the reliability and consistency with other different prediction techniques already proposed: for the next solar cycle our approach confirms the actual trend of a progressive reduced solar activity.

\end{document}